\documentclass[sigconf, nonacm]{acmart}

\usepackage{enumitem}
\usepackage{url}
\usepackage{tabularx}
\usepackage{tablefootnote}
\usepackage{tikz}
\usepackage{listings}

\usepackage{footnote}
\usepackage{algorithm, setspace}
\usepackage{multirow}
\usepackage{caption}
\usepackage{subcaption}
\usepackage[noend]{algpseudocode}
\floatname{algorithm}{Algorithm}

\definecolor{dkgreen}{rgb}{0,0.6,0}
\definecolor{gray}{rgb}{0.5,0.5,0.5}
\definecolor{mauve}{rgb}{0.58,0,0.82}
\definecolor{revision}{RGB}{3, 140, 252}
\def\ours{{LaPuda}}

\newcommand\vldbdoi{XX.XX/XXX.XX}
\newcommand\vldbpages{XXX-XXX}
\newcommand\vldbvolume{14}
\newcommand\vldbissue{1}
\newcommand\vldbyear{2020}
\newcommand\vldbauthors{\authors}
\newcommand\vldbtitle{\shorttitle} 

\newcommand\vldbpagestyle{plain}

\begin{document}
\title{LaPuda: LLM-Enabled Policy-Based Query Optimizer for Multi-modal Data}

\author{Yifan Wang $^*$}
\affiliation{%
  \institution{University of Hawaii}
}
\email{yifanw@hawaii.edu}

\author{Haodi Ma $^*$}
\affiliation{%
  \institution{University of Florida}
}
\email{ma.haodi@ufl.edu}

\author{Daisy Zhe Wang}
\affiliation{%
  \institution{University of Florida}
}
\email{daisyw@ufl.edu}

\begin{abstract}
Large language model (LLM) has marked a pivotal moment in the field of machine learning and deep learning. Recently its capability for query planning has been investigated, including both single-modal and multi-modal queries. However, there is no work on the query optimization capability of LLM. As a critical (or could even be the most important) step that significantly impacts the execution performance of the query plan, such analysis and attempts should not be missed. 

From another aspect, existing query optimizers are usually rule-based or rule-based + cost-based, i.e., they are dependent on manually created rules to complete the query plan rewrite/transformation. Given the fact that modern optimizers include hundreds to thousands of rules, designing a multi-modal query optimizer following a similar way is significantly time-consuming since we will have to enumerate as many multi-modal optimization rules as possible, which has not been well addressed today.

In this paper, we investigate the query optimization ability of LLM and use LLM to design \textbf{LaPuda}, a novel \textbf{L}LM \textbf{a}nd \textbf{P}olicy based m\textbf{u}lti-mo\textbf{da}l query optimizer. Instead of enumerating specific and detailed rules, LaPuda only needs a few abstract \textit{policies} to guide LLM in the optimization, by which much time and human effort are saved. Furthermore, to prevent LLM from making mistakes or negative optimization, we borrow the idea of gradient descent and propose a \textit{guided cost descent} (GCD) algorithm to perform the optimization, such that the optimization can be kept in the correct direction. In our evaluation, our methods consistently outperform the baselines in most cases. For example, the optimized plans generated by our methods result in 1$\sim$3x higher execution speed than those by the baselines.           

\end{abstract}

\maketitle

\pagestyle{\vldbpagestyle}
\begingroup\small\noindent\raggedright\textbf{PVLDB Reference Format:}\\
\vldbauthors. \vldbtitle. PVLDB, \vldbvolume(\vldbissue): \vldbpages, \vldbyear.\\
\href{https://doi.org/\vldbdoi}{doi:\vldbdoi}
\endgroup
\begingroup
\renewcommand\thefootnote{}\footnote{\noindent
This work is licensed under the Creative Commons BY-NC-ND 4.0 International License. Visit \url{https://creativecommons.org/licenses/by-nc-nd/4.0/} to view a copy of this license. For any use beyond those covered by this license, obtain permission by emailing \href{mailto:info@vldb.org}{info@vldb.org}. Copyright is held by the owner/author(s). Publication rights licensed to the VLDB Endowment. \\
\raggedright Proceedings of the VLDB Endowment, Vol. \vldbvolume, No. \vldbissue\ %
ISSN 2150-8097. \\
\href{https://doi.org/\vldbdoi}{doi:\vldbdoi} \\
}\addtocounter{footnote}{-1}\endgroup
\def\thefootnote{*}\footnotetext{These authors contributed equally to this work.}\def\thefootnote{\arabic{footnote}}

\section{Introduction}
\label{sec:introduction}
In today's rapidly advancing technological landscape, the emergence of large language model (LLM) has captivated the attention of researchers. LLM has the capacity to comprehend and generate human-like text at an unprecedented scale, and they have set new benchmarks in a range of applications, including text generation~\cite{vaswani2017attention, colas2023can}, translation~\cite{devlin2018bert, floridi2020gpt}, sentiment analysis~\cite{zhang2023instruct, wang2023chatgpt}, and conversational AI~\cite{floridi2020gpt, chen2023shikra}. By training on vast amounts of data, these models possess an extraordinary capacity to understand context, parse nuanced meanings, and generate coherent and contextually appropriate responses. Furthermore, LLM has developed the capability to handle other data modalities in addition to text, like image, audio and video. Recently multi-modal LLM is becoming increasingly frequently used in various applications. 

Given the powerful data understanding and reasoning abilities of LLM, researchers are exploring the potential of LLM as planner for human-language-described tasks and queries. Chain-of-Thought (CoT)~\cite{wei2022chain}, as a representative, provides reasoning examples to LLM to guide it for similar planning tasks. Other recent methods~\cite{besta2023graph, yao2023tree, hao2023reasoning} focus on leveraging different searching algorithms or structures, e.g. tree search, to enhance LLM planning ability. Other works~\cite{zhou2022least, welleck2022generating, Shinn2023ReflexionAA, paul2023refiner, wang2022self} propose further improvement via task decomposition, feedback from LLM, or aggregation on reasoning paths. 

Most of the aforementioned works are for single-modal queries, involving only text data. With the recent trend of multi-modal query processing, multi-modal query planning based on LLM has been initially studied\cite{urban2023caesura, reimage-rag-multimodal}. But for real-world use cases where the queries are complex, getting the optimal query plan is more important than just generating a valid plan. Existing LLM-planning approaches suffer from the following problems in terms of query optimization: (1) few of them provide concrete study on the further optimization for the generated plan, instead, they usually only focus on how to let LLM generate an executable plan given the query, without considering the plan execution efficiency. And (2) even though they can be further developed to optimize the generated plan, the optimization will heavily rely on human involvement, making the optimization difficult to proceed. For example, least-to-most prompting requires manually crafting specific and detailed examples for task decomposition and sub-task solving. Such examples are hard to get in multi-modal scenarios, considering the wide range of possible operations and optimizations. As a result, the lack of optimization makes the existing planning methods impractical to be applied in the real world, as the execution of the non-optimal plan results in unnecessarily high latency.

In this paper, beyond its traditional application in planning, we explore LLM as a viable solution for formulating a multi-modal query optimizer. Specifically, the central challenges in the design of a multi-modal query optimizer include (1) the enumeration of multi-modal query rewrite and transformation rules, (2) the semantic transformation for multi-modal query plan, and (3) the cost model for multi-modal query plan. It is imperative to underscore that all three dimensions remain inadequately addressed and explored. The first dimension requires rich experience and a substantial temporal commitment, given that a mature optimizer conventionally entails hundreds to thousands of rules. The second challenge is also hard to address as traditional rules have no way to handle semantically (but not syntactically) equivalent plan transformation. For the third challenge, a mature cost model has to consider many factors including IO, CPU and memory access, network speed, etc, which heavily relies on human expertise. And even with rich human experience, making the cost model accurate is still an open question until today. 

LLM is suitable to tackle the first and second dimensions of the challenges. Traditional rule-based query optimizer has to follow detailed and explicit rules to transform the query plan. In addition, those rules are usually structure-based (i.e., based on sub-tree structures of the plan) which cannot handle semantic operations like visual QA and text QA that rely on question semantics. Thanks to the capability of LLM to understand high-level human language and concepts, it can effectively transform the plan (1) both structurally and semantically, (2) simply based on abstract, higher-level instructions, which we refer to as \textit{policy}, instead of detailed rules. Compared to a rule, a policy is a summary of multiple transformation actions. More details of our policies can be found in section~\ref{sec:Policy}.

However, negative optimization is unavoidable in the optimization purely using LLM, i.e., the optimized plan may have a longer execution time than the initial plan. 
To fill in this gap, we have to design our own cost model to detect negative optimization. 
To reduce the dependence on human experience (as mentioned in the third challenge above), our cost model considers much fewer factors. Experiments show such a roughly designed cost model is effective enough to guide LLM for high-quality optimization. 
In addition, we evaluate the possibility of using LLM itself as a cost model, which reveals its limitation in such a task at the current stage, as well as its potential for the future. More details about this are discussed in Section~\ref{sec:llm-cost-model}.   

In this paper, we propose \textbf{LaPuda}, a novel \textbf{L}LM \textbf{a}nd \textbf{P}olicy based m\textbf{u}lti-mo\textbf{da}l query optimizer. LaPuda consists of a pre-trained LLM, a few high-level and abstract policies, and a supervisor including error monitor, cost model and other auxiliary components.  
The capability of using abstract policies saves much time from enumerating detailed and complex rules. Moreover, leveraging the principle of gradient descent in deep learning, we propose a novel \textit{guided cost descent} (GCD) algorithm to conduct the optimization procedure. The core concept revolves around ensuring the optimization remains on the correct trajectory using cost estimation and corresponding feedback to prevent LLM from generating invalid plans and negative optimization. 
The technical details are included in Section~\ref{sec:lapuda}, \ref{sec:op-and-policy} and \ref{sec:cost-descent}. 

This paper demonstrates that LLM can adeptly comprehend policies and generate appropriately optimized plans. Furthermore, LLM capitalizes on its pre-existing knowledge of traditional optimizers, thereby economizing on design costs. Our experiments indicate that providing examples solely for non-SQL operators suffices for LLM, obviating the need for specific instances for SQL operators, as such knowledge is already preserved by the pre-trained model. 

We conduct evaluations based on the image-relational scenario, which handles the query operators for relational and image data. The evaluation shows that our methods achieves higher effectiveness on optimization than the state-of-the-art LLM reasoning methods. The resulting plans generated by our methods are executed significantly faster than those by the baselines. Compared to the most effective baseline method, the plans optimized by our methods have up to 2x average execution speed.

To our best knowledge, \ours\ is among the first optimizers for multi-modal queries, as well as the first attempts to build a query optimizer using LLM to be facilitated by its tremendous scale of possessed knowledge and powerful reasoning capability.  

The main contributions of this paper are shown below:
\begin{enumerate}[leftmargin=2em]
\item We build LaPuda, the first policy-based multi-modal query optimizer using LLM, which no longer needs optimization rules and thus takes the first step of designing new types of query optimizers in the era of LLM. It is also among the first studies to explore the query-optimizing capability of LLM.
\item We propose a novel guided cost descent (GCD) mechanism to guide the optimization process such that it can keep on the correct direction, which effectively improves the optimization quality.
\item We propose an effective two-level guidance strategy to provide feedback and guide LLM to handle errors and locate better direction during the optimization.   
\item We conduct extensive evaluations for the performance of LaPuda and its variants, as well as evaluation for the impact of critical components in our methods, which affirms the effectiveness of our ideas and design.  
\end{enumerate}

This paper is organized as follows: Section~\ref{sec:related-work} introduces works related to LaPuda. Section~\ref{sec:lapuda} introduces the architecture and workflow of LaPuda. The following Sections \ref{sec:op-and-policy} and \ref{sec:cost-descent} present technical details for the major components. Section \ref{sec:llm-cost-model} introduces our attempt to leverage LLM as our cost model. Finally, Section~\ref{sec:exp} presents experiments to evaluate the performance of our methods and the effect of the core components.   

\begin{figure*}
  \centering
  \includegraphics[width=\textwidth]{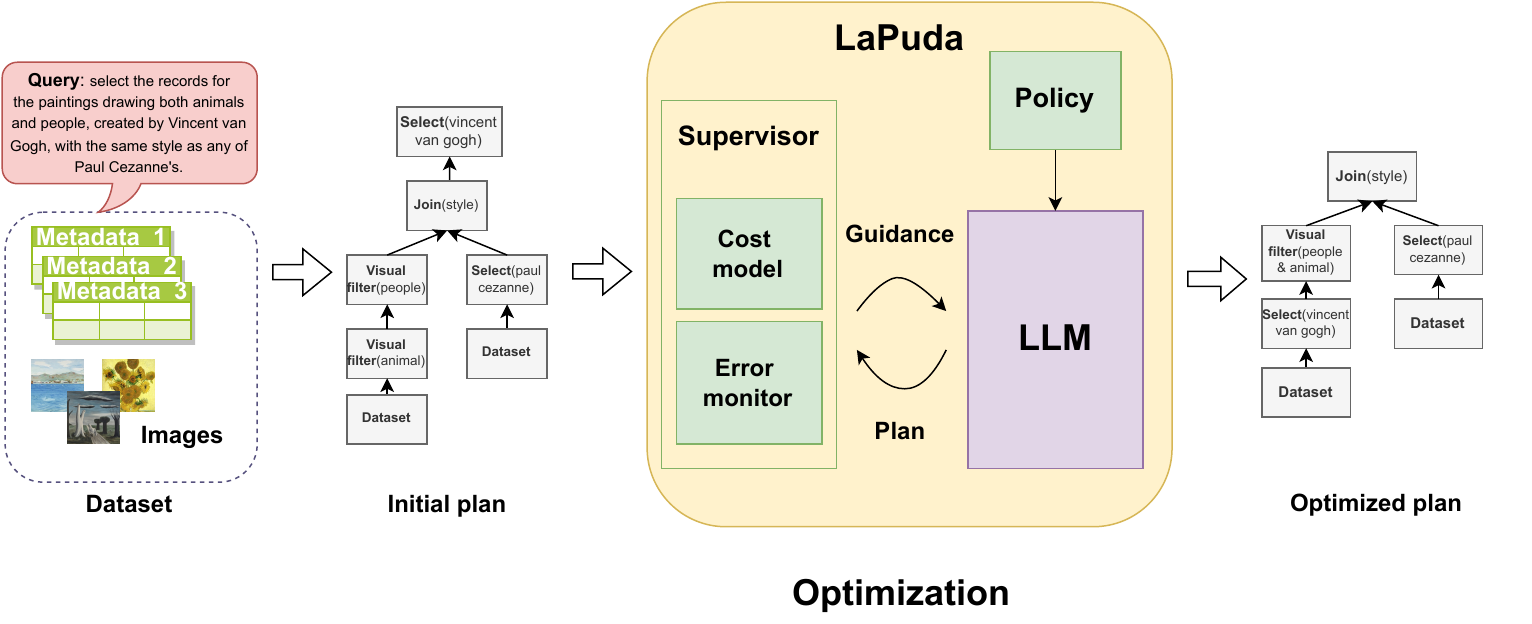}
  \caption{Architecture and workflow of \ours}
  \setlength{\belowcaptionskip}{-50pt}
  \label{fig:arch}
\end{figure*}

\section{Related work}
\label{sec:related-work}
LLM reasoning typically involves breaking down intricate queries into sequential intermediate steps, i.e. chains, prior to generating the final response. It is similar to the planning ability in intelligent agents, which requires generating a series of actions to achieve a specific goal~\cite{mccarthy1963situations, bylander1994computational}. Such ability is exemplified by the Chain-of-Thought (CoT) prompting and its variations~\cite{wei2022chain, kojima2022large}. Since CoT generates all steps simultaneously which may lead to error propagation, Self-Consistency~\cite{wang2022self} employs the sampling of multiple chains and selects the best answer through majority voting. Least-to-most prompting~\cite{zhou2022least} simplifies the task into simple sub-tasks and solves them sequentially. However, such methods are limited to tasks with fixed solving patterns, while multi-modal query optimization requires intertwined instead of sequential reasoning. Other works, including self-evaluation approaches, either explore providing feedback for intermediate steps~\cite{welleck2022generating, Shinn2023ReflexionAA, paul2023refiner}, or utilize heuristic-based search techniques such as depth-/breadth-first search~\cite{yao2023tree} or Monte Carlo
Tree Search (MCTS) for improved reasoning paths. LaPuda shares a similar spirit to these works. However, LaPuda leverages general policies in natural language and requires less explicit examples to demonstrate all the rules or reasoning steps, increasing applicability in tasks that lack human knowledge like multi-modal query optimization. 

Besides, Classical planning methods have been widely adopted in robots and embodied environments~\cite{camacho2013model, jiang2019task}. Recently, prompting LLMs to do planning directly has gained attention and shown potential~\cite{huang2022inner, singh2022progprompt, ding2023task}. Moreover, based on LLMs' powerful programming ability~\cite{lyu2023faithful, jojic2023gpt, liu2023llm+}, recent works first translate natural language instructions into the executable programming languages, such as Planning Domain Description Language (PDDL), and runs classical planning algorithms, such as LLM+P~\cite{liu2023llm+}. CAESURA~\cite{urban2023caesura} goes outside the scope of code generation to generate reasoning plans for natural language multi-modal queries. However, to the best of our knowledge, no system has been proposed yet that leverages the reasoning capabilities of LLMs to optimize complex query plans for multi-modal data lakes, where the efficiency of plans is one of the main concerns.

\section{LaPuda}
\label{sec:lapuda}
In this section, we introduce the overall architecture and workflow of \ours. Then in the following sections, we discuss technical details about the major components in \ours. 

\label{sec:arch}
Figure~\ref{fig:arch} illustrates the architecture and overall workflow of \ours. As a query optimizer, \ours\ works on optimizing the initial plan and generating the finally optimized plan. The initial plan is generated using any existing multi-modal query plan generator (like \cite{urban2023caesura}) which converts the raw natural language query into an operation tree without further optimization. For example, in Figure~\ref{fig:arch}, the query requires selecting the records of paintings based on artist and painting style (which need relational operators in RDBMS), as well as the subject conditions (which require visual operators like object detection). The initial plan accordingly consists of standard SQL join and select, and visual filters (based on object detection techniques). The initial plan is usually non-optimal and will be fed to \ours\ for further optimization. Then the resulting optimized plan will be executed by the following executor. 

Shown as the yellow block in Figure~\ref{fig:arch}, \ours\ has three major components: (1) LLM (purple box), (2) supervisor (green border rectangle), and (3) policy (green box). The LLM can be any state-of-the-art LLM, which makes \ours\ a generic framework. The policy module includes a set of policies, where each is an abstract, high-level instruction that summarizes a family of optimization rules, covering operators in various modalities. We will introduce the details for the policy in Section~\ref{sec:op-and-policy}. The supervisor plays a core role in \ours, which interacts with the LLM to iteratively receive the newly generated plan and guide the LLM to further optimize the latest plan or fix errors found in the plan. The supervisor consists of two critical modules: a cost model to estimate the execution cost of the received plan, and an error monitor to check for invalid plans (with structural errors) and inequivalent plans (which are not equivalent to the initial plan). We also design a two-level guidance strategy to guide LLM during the optimization: The error monitor provides coarse-level guidance and a finer-level guidance on the optimization direction is conducted based on the estimated cost from the cost model. Furthermore, we propose an algorithm utilizing all those components and guidance to effectively drive LLM to optimize the initial plan. The algorithm is one of the major novelties in this paper, called ``guided cost descent'' (GCD). Essentially, guided cost descent effectively enhances the optimization by following the correct direction provided by the supervisor based on estimated cost, and also reduces negative optimization during the process. We discuss GCD in detail in Section~\ref{sec:cost-descent}.

Such an architecture of \ours{} (1) enables effective multi-modal query optimization without specifying detailed and explicit rules like in traditional rule-based optimizer, which saves significant development on enumerating rules, and (2) provides guidance to keep LLM in the correct direction, which reduces the negative optimization, effectively deepens the optimization and achieves higher-quality optimized plans.

\section{Operator and policy}
\label{sec:op-and-policy}
In this section, we discuss the operators we focus on and the policies we are currently using in this work.

\subsection{Operator}

Given our focus on multi-modal query optimization, the query operators encompass both conventional SQL operators and those related to visual elements. We only consider the conjunctive cases of multiple operators and exclude the union cases from this study, i.e., there could be ``AND'' logic but no ``OR'' in our queries.    

\noindent\textbf{SQL operators:} To streamline the considerations, we limit our analysis to the two most prevalent SQL operators: \texttt{Select} and \texttt{Join}. As illustrated in Code \ref{code:SQL_operators}, the \texttt{Select} operator takes a table with a designated column and specified condition as input. A simple condition consists of the table and column names, a comparison operator, and a corresponding value. A complex condition includes the conjunction of multiple simple conditions (i.e., clauses connected by the \texttt{AND} logical operator). The \texttt{Join} operator involves two tables and the joining columns. For simplicity, we restrict our consideration to equal-join, where the keys being joined are equal.
\begin{lstlisting}[language=Python, caption=SQL operators, label=code:SQL_operators]
## Select
Select(Table.column <comp> <value> [AND ...])

## Join
Join(Table.column = Table.column)
\end{lstlisting}

Our objective is not to supplant the conventional query optimizer with LaPuda. Instead, our emphasis lies in introducing a pivotal component for an end-to-end LLM-based query engine, specifically, LLM-based query optimizer. Additionally, we aim to unveil the prospect of leveraging LLM for the construction of a multi-modal query optimizer, a venture that traditional optimization techniques have not undertaken. Therefore, our scope does not encompass coverage of all the operators in relational algebra but concentrates on \textit{select} and \textit{join} operations.

\noindent\textbf{Visual-related operators:} We examine two common operators related to visual elements (e.g. images): object detection and object counting. The \texttt{Object detection} operator assesses whether the specified object is present in the given image(s) by natural language question and outputs the rows where the result is \texttt{True}. In our implementation, it operates on the column of image paths, and the visual component takes the corresponding image(s) as input, performs object detection, and produces a \texttt{True/False} outcome for each image.

Similarly, the \texttt{Object counting} operator checks for the presence of multiple target objects in a given image. In addition to the table, column, and question about the target object, this operator requires a threshold (i.e., target number). It conveys all images to the visual component based on their paths in metadata, and returns metadata rows where the corresponding image includes more than threshold target objects. The structures of these two operators within our queries are demonstrated in Code \ref{code:visual_operators}.

\begin{lstlisting}[language=Python, caption=Visual-related operators, label=code:visual_operators]
## Object detection
Object detection(Table.column: are there XXX [AND YYY [AND ...]]?)

## Object counting
Object counting(Table.column: how many XXX are there?: <threshold>) 
\end{lstlisting}

\subsection{Policy}
\label{sec:Policy}
In this section, we will discuss the three categories of policies we introduce to the large language model. We represent each query plan as a binary tree of operators, and the policies aim at transforming the tree structure.  

\subsubsection{Operator movement}\hfill\\
Generally speaking, in a multi-modal query, we prefer to apply the more expensive operation later in the execution, in which case it will deal with less amount of data (except in some cases for \texttt{Join}). For example, based on the order of the operator costs (shown in Code 3), \texttt{Object detection} should be executed later than \texttt{Select} as \texttt{Select} can reduce the number of rows passed through \texttt{Object detection}. Thus, Policy 1, the \textit{operator movement policy}, is essentially a couple of instructions teaching LLM that it is beneficial to move more expensive operators closer to the root of the plan tree.

\begin{lstlisting}[language=Python, caption=An example for Policy 3, label=code:Policy]
## Operator order
(cheapest) Select < Join < Object detection < Object counting (most expensive)

'''
Example for Policy 3
'''
initial_plan = { "Operator": Object Counting(table_3.col_3: how many men are there?: <threshold>),
    "Left_child": {
        "Operator": Object Detection(table_3.col_3: is there any man?)
        "Left_child": Table_3,
        "Right_child": None,
    }, 
    "Right_child": None
}
optimized_plan = { "Operator": Object Counting(table_3.col_3: how many men are there?: <threshold>),
    "Left_child": Table_3,
    "Right_child": None
}
\end{lstlisting}

\subsubsection{Operator merge}\hfill\\
When multiple same-type filters (e.g., two \texttt{Select} or two \texttt{Object detection}) are applied to identical target, an intuitive optimization involves consolidating them into a single operation. For example, when two \texttt{Select} operators target the same column, e.g. both on \texttt{Table\_1.col\_1}, they can be merged into one \texttt{Select} operator with the conjunction of their conditions. Analogously, when two \texttt{Object detection} (e.g., \texttt{Are there men?} and \texttt{Are there women?}) are sequentially performed on the same set of images, consolidation is viable (which results in \texttt{Are there both women and men?}). This optimization strategy streamlines the plan execution by minimizing redundant iterations over identical data, thereby enhancing efficiency compared to applying the operations separately.

\subsubsection{Operator removal}\hfill\\
Taking a step beyond operator merging, when multiple filters of the same type are applied to an identical dataset, there may be instances where one filter's condition is more restrictive than another's. In such cases, the operation with the less stringent condition can be removed. For example, in the scenario where two object counting operations are executed on the same image set, both targeting the same object (like both asking \texttt{How many dogs}), removing the operation with the smaller threshold accomplishes the same outcome. 

It is important to note that when an \texttt{Object detection} and an \texttt{Object counting} operation are executed on the same image set and pertain to the same object, the object detection operation can be omitted. This is because following the definition, the object counting operation filters results based on the non-negative threshold, ensuring that its outcomes satisfy the conditions of the object detection operation. For instance, in the example for Policy 3 in Code \ref{code:Policy}, the initial plan contains \texttt{Object counting} and \texttt{Object detection}, both on \texttt{table\_3.col\_3}. According to Policy 3, the \texttt{Object detection} operation can be removed while retaining only \texttt{Object counting} is a valid optimization.

To summarize, the optimization policies we currently use in \ours\ are listed as follows:
\begin{enumerate}
\item[1. ]\textbf{Operator movement}: Expensive operators should be closer to the root while cheaper operators should be further from the root in the plan.
\item[2. ]\textbf{Operator merge}: Same-type filters on the same target can be merged when they are direct parents and children.
\item[3. ]\textbf{Operator removal}: For operations applied on the same target, the ones with the looser conditions can be removed when they are direct parents or children to the stronger ones.
\end{enumerate}
Each policy covers several different rules involving different operators. Compared to those detailed rules, a policy is more abstract and can be stated using only one or two sentences in the prompt, which significantly saves human effort in designing the optimizer.   
Note that all our policies are based on the fact that an initial binary-tree structured plan has been given. We do not discuss how to parse a query and generate the initial plan in this paper, since there has been work on this topic~\cite{urban2023caesura}.

\section{Supervisor and guided cost descent}
\label{sec:cost-descent}
We borrow the idea of gradient decent and design \textit{guided cost descent} (GCD) to effectively guide LLM optimizing in the desired direction, where ``guided'' means the descent process is guided by an external supervisor, unlike gradient descent that happens inside the learning model. 

Guided cost descent is essentially a two-level guidance-based strategy for enhancing LLM-based query optimization. The strategy provides a coarse-level and a finer-level guidance to present the correct optimization direction to LLM. The coarse guidance is provided when errors are found in the generated plan (in which case we call the plan an \textit{invalid plan}). Such guidance will point out that errors are existing in the generated plan and let LLM re-generate the plan based on the last valid plan while avoiding previous errors. 
The finer guidance is based on cost estimation for the initial and generated plans. It will give LLM corresponding feedback based on the change in plan cost. This is called finer since it provides clearer information about how good or bad the current direction is. 
In this section, we introduce more technical details for GCD.

\begin{figure}
  \centering
  \includegraphics[width=\columnwidth]{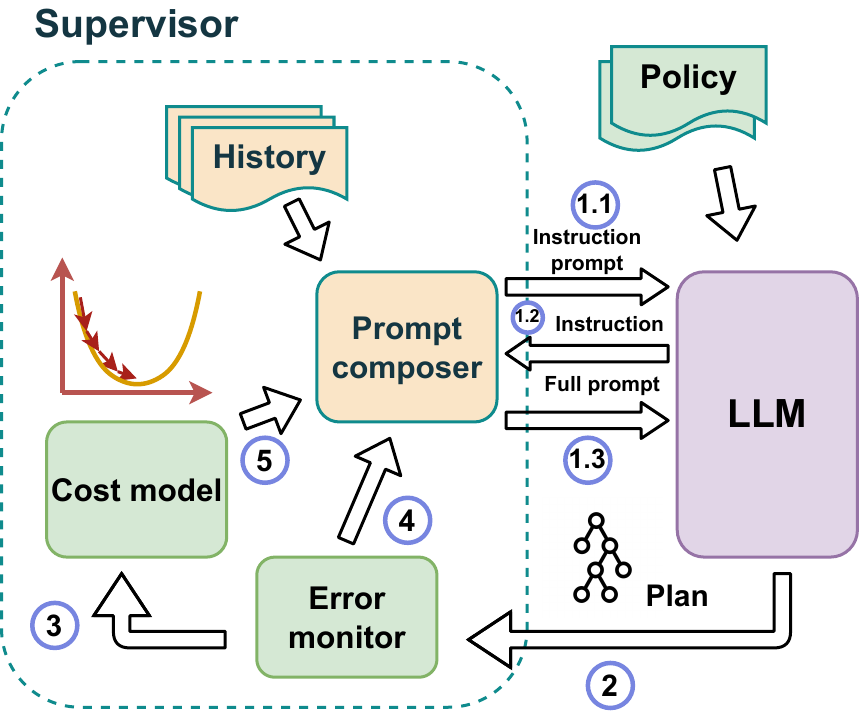}
  \caption{Guided cost descent workflow}
  \setlength{\belowcaptionskip}{-50pt}
  \label{fig:gcd}
\end{figure}

\subsection{Components and workflow}
Figure~\ref{fig:gcd} zooms in the supervisor and illustrates its components in detail. The major components are the cost model and error monitor, where the former computes the cost for each plan and the latter checks whether the optimized plan is valid. In this paper, an invalid plan is a plan having errors, which could be either structural errors (like a join operator only having one child or a selection predicate referring to a table not existing in its sub-tree) or inequivalent to the initial plan. More details can be found in Section~\ref{sec:correct-and-recover}. 
There are also two minor components: the history management module and the prompt composer, respectively working for maintaining the history information (like generated plans and their costs) and composing the prompt to deliver the latest guidance to LLM.  

Algorithm~\ref{alg:gcd} presents the detailed workflow of the end-to-end GCD process.
For simplification, in the following discussion, "Arrow X" means the arrow marked with the number X in Figure~\ref{fig:gcd}, while ``line Y'' refers to the Y-th line in Algorithm~\ref{alg:gcd}.

\begin{enumerate}[leftmargin=2em]
    \item \textbf{Initialization (line 1-6)}: At first, supervisor 
    sets the initial variables, including setting the initial plan, computing its cost, initializing the tracker variables like best plan until now, latest valid plan and cost, etc. 
    The supervisor also sets an internal termination tolerance $t$ remarking the maximum number of consecutive invalid or non-improved plan generations that we allow before terminating GCD.  
    \item \textbf{One-step optimization (line 8)}: 
    This step is shown as Arrows 1.1, 1.2, 1.3, and 2 in Figure~\ref{fig:gcd}, corresponding to the procedure $\operatorname{GetPlan}$ in Algorithm~\ref{alg:gcd}. 
    The supervisor first uses ``instruction prompt'' to let LLM propose how it will instruct itself to optimize the plan (Arrow 1.1), and the returned instruction (Arrow 1.2) will be combined with all other necessary information (other instructions, policies, history, etc.) as the full prompt to LLM which starts the plan generation (Arrow 1.3).   
    Then LLM generates and returns an optimized plan to the supervisor (Arrow 2). 
    \item \textbf{Error handling (line 9-10, 24-26)}: The error monitor determines if the optimized plan is valid by detecting any structural errors in it and checking its equivalence to the initial plan, which is the $\operatorname{CheckError}$ procedure (line 9). If the plan is valid, it will be passed to the cost model (line 11, Arrow 3) for estimating the plan cost; otherwise, the cost estimation will be skipped (line 24) and a corresponding \textit{feedback} (line 26) becomes part of the prompt (Arrow 4) for the next iteration to prevent LLM from producing the same errors in future optimization. 
    \item \textbf{Cost estimation (line 11-23)}: The cost model works on estimating the cost for a given plan (i.e., $\operatorname{GetCost}$ in line 11). If the current valid plan has a lower cost than the last valid plan 
    , it is a correct optimization. Otherwise, it will be also counted as a wrong optimization (like the case of an invalid plan) and make the total count further approach the tolerance (lines 12-14). The cost will then become part of the feedback to LLM (Arrow 5, line 14/20). Moreover, if the current cost is lower than the best plan cost, the best plan will be updated accordingly (lines 16-18). 
    \item \textbf{Prompt preparation for next iteration}: Several variables will be updated for constructing the next prompt: the feedback will include the current plan cost and judgment on whether the plan is improved or not (line 14/20), the latest valid plan and cost are updated accordingly (line 21), then they are also added into the history (line 22-23). Note that the latest plan and cost, as well as the history, will not be updated if the current plan is invalid, i.e., the history only includes valid plans.  
    \item \textbf{Next plan generation or termination}: GCD will repeat the steps above from ``one-step optimization'' to ``prompt preparation for next iteration'' until the number of consecutive wrong optimizations reaches the tolerance. In each new iteration, the LLM receives the policies, examples, history, and feedback, as well as the latest valid optimized plan, to generate the next optimized plan.  
\end{enumerate}

\begin{algorithm}
\setstretch{1.2}
\caption{Guided cost descent}
\label{alg:gcd}
\begin{algorithmic}[1]
\Require{Policies $\mathcal{R}$, example initial-optimized plan pairs $\mathcal{P}_e$, target initial plan to be optimized $p_0$, termination tolerance $t$} 
\Ensure{The best optimized plan $p^\ast$}
\State initial plan cost $c_0 := \operatorname{GetCost}(p_0)$
\State latest valid plan cost $c := c_0$, latest valid plan $p := p_0$ 
\State best plan until now $p^\ast := p_0$, best cost $c^\ast := c_0$
\State history optimized plans $\mathcal{P}_h := \emptyset$, corresponding history plan costs $\mathcal{C}_h := \emptyset$
\State feedback $f$ := ``''
\State number of wrong optimization $n_w := 0$ 
\While{$n_w < t$} \Comment{stopping condition not satisfied}
    \State optimized plan $p_i := \operatorname{GetPlan}(\mathcal{R}, \mathcal{P}_e, \mathcal{P}_h, \mathcal{C}_h, c, p, f)$
    \State plan errors $\mathcal{E}_i := \operatorname{CheckError(p_i)}$
    \If{$\mathcal{E}_i = \emptyset$} \Comment{$p_i$ is valid}
        \State current plan cost $c_i := \operatorname{GetCost}(p_i)$
        \If{$c_i \geq c$} \Comment{cost is increased}
            \State $n_w := n_w + 1$
            \State $f$ := Concat(``No improvement'', $c_i$)
        \Else
            \If{$c_i < c^\ast$}
                \State $p^\ast := p_i$ \Comment{update the best plan}
                \State $c^\ast := c_i$
            \EndIf
            \State $n_w := 0$
            \State $f$ := Concat(``Improved'', $c_i$)
        \EndIf
        \State $c := c_i$, $p := p_i$
        \State $\mathcal{P}_h := \mathcal{P}_h \cup \{p\}$
        \State $\mathcal{C}_h := \mathcal{C}_h \cup \{c\}$
    \Else \Comment{$p_i$ is invalid}
        \State $n_w := n_w + 1$ 
        \State $f$ := ``No valid optimization generated''
        
    \EndIf
\EndWhile
\State \textbf{return} $p^\ast$
\end{algorithmic}
\end{algorithm}

\subsection{Coarse-level guidance}
\label{sec:correct-and-recover}
\subsubsection{Error detection}\hfill\\
We categorize the errors in invalid plans into two classes: structural error and in-equivalence. Structural error refers to any error in the plan tree structure, for example, children-number errors (like a join having only one child). In-equivalence means the optimized plan is structurally correct but not equivalent to the initial plan. 

Accordingly, we design the error monitor to detect those two classes of errors. For structural error, we manually design a structure and syntax inspection module to check the error based on sub-tree pattern matching, i.e., it attempts to match sub-tree structures with pre-defined error structure patterns and reports the error when there is a match. 

Detecting in-equivalence is harder than structural error since the equivalence in multi-modal queries includes both syntax and semantic equivalence. For example, in a visual QA operator, ``how many persons'' and ``how many people'' are semantically equivalent, which cannot be figured out by purely structural inspection. To cover those cases, we design a hybrid in-equivalence inspector that combines structural inspection (based on standard relational algebra) and semantic matching based on machine learning. Specifically, the inspector jointly uses a BERT-based pre-trained model and TF-IDF to get a comprehensive semantic similarity for the two natural language sentences being compared. The individual similarity scores from BERT and TF-IDF are both defined as cosine similarity between the vector representations (either TF-IDF vectors or BERT embeddings) of the two sentences. Then the two scores are aggregated and thresholded to determine whether the two sentences are equivalent or not. In our evaluation, the aggregation is the multiplication of them, and the threshold for final determination is 0.5.

The reason for not purely relying on the BERT-based model is that the model fails to distinguish some instances, like ``men'' and ``women'' usually result in a high similarity score by the model. Therefore, TF-IDF is added to enable auxiliary term-based similarity matching. 

\subsubsection{Recovery from error}\hfill\\
\label{sec:err-recover}
When the current plan is invalid, the LLM needs to explore other possible optimization directions and avoid the existing errors, which we call "recovery" from the error. 
Specifically, the supervisor passes the history until the latest valid plan and the feedback message indicating that no valid optimized plan was generated, along with other necessary information (listed as the parameters of $\operatorname{GetPlan}$), to LLM, letting LLM re-generate the optimized plan starting from the latest valid plan while avoiding the same errors. If LLM keeps outputting invalid plans, GCD will terminate and return the best plan during the optimization.

In summary, such error-driven guidance cannot give more information other than that the current optimization is wrong, which is relatively coarse. But being coarse does not mean being ineffective. Instead, compared with passing a large amount of context to LLM, a simple and strong order is more compelling, due to the long-term memory issue of LLM~\cite{llm-long-term-mem-zhong2023memorybank}.

\subsection{Finer-level guidance}
\label{sec:cost-model}
\subsubsection{Cost model}\hfill\\
In addition to the coarse guidance, \ours{} also provides a cost-based finer-level guidance that informs more details to LLM and leads to a finer tuning on the optimization direction. \ours\ has integrated a simple cost model which is proved to be effective in the evaluation in Section~\ref{sec:exp-cost-model}.            

Specifically, the cost model estimates the plan cost according to Equation~\ref{eq:cost}, 
\begin{align}
\label{eq:cost}
    c_r = {\rho}_{r}\sum_{i \in S_{r}} N_{i}
\end{align}
where $c_r$ is the cost of the current plan tree rooted at operator $r$, ${\rho}_{r}$ is the cost factor for $r$, i.e., the estimated cost to process each row passed though $r$, $S_r$ is set of the child operator(s) of $r$, while $N_i$ is the number of rows output by the child operator $i$. Then the number of rows output by $r$ is estimated by Equation~\ref{eq:num-rows},  

\begin{align}
\label{eq:num-rows}
    N_r = {\alpha}_{r}\sum_{i \in S_{r}} N_{i}
\end{align}
where ${\alpha}_{r}$ is the selectivity of $r$, i.e., the ratio of output rows over input rows. Note that both ${\rho}_{r}$ and ${\alpha}_{r}$ are roughly estimated in the cost model without utilizing detailed database statistics except the number of rows in each table.   

\subsubsection{Guidance composition}\hfill\\
In the cases where errors are found (i.e., coarse-level guidance), although the current direction is known to be wrong and must be changed, the correct direction is still unclear. 
Therefore, we design the finer-level guidance to further utilize the power of LLM itself for reasoning and finding the correct direction. 

The cost model will estimate the cost for each new valid plan. 
Then based on the costs of the current and the previous plans, the guidance prompt consists of extensive information, including the corresponding feedback based on whether the plan is improved, the history of valid plans and their costs, as well as the policies, examples and the current plan to be further optimized. Such cost-based guidance provides a more complete view for LLM to understand the history experience, the current situation, and at what stance each previous step optimizes the initial plan, which is essential to LLM to fine-tune the optimization direction by itself.

\subsection{Feedback}
As introduced above, there are two types of feedback, one is for reporting errors (line 26 in Algorithm~\ref{alg:gcd}) while the other is for indicting the cost change (line 14/20 in Algorithm~\ref{alg:gcd}). We denote the former by \textit{error-feedback} and the latter by \textit{cost-feedback}. Particularly, error-feedback is essential to the coarse-level guidance while cost-feedback plays a critical role in finer-level guidance. 

To prove the necessity of the two-level guidance strategy, we implement a coarse-level-only version of \ours, named \textit{\ours-lite}. \ours-lite still uses the same supervisor as \ours, and still reports errors to LLM for coarse-level guidance. The only difference is that \ours-lite never provides information about plan cost and whether the plan has been improved to LLM, i.e., \ours-lite does not have cost-feedback, thus also has no finer-level guidance. In Section~\ref{sec:exp}, we show that \ours \ outperforms \ours-lite in most cases.

\section{LLM as cost model}
\label{sec:llm_cost_model}
LLM is not good at quantitative/numerical reasoning~\cite{llm-quantitive-reasoning-1, llm-quantitive-reasoning-2}, which makes it hard to be directly adopted as the cost model. However, we find another way to use LLM as a cost model: Since the key to cost-feedback is the determination of whether the plan has been improved, we may use LLM as a classifier that makes the determination based on the plan structures and basic database statistics to avoid complex calculation. Note that the task of cost-feedback can only be ``roughly'' treated as a classification since the concrete cost value also plays a unique role in the optimization to guide LLM to choose the likely most significant direction. We will further explore the power of the cost value in cost feedback in our future work. In this paper, we just simplify the problem and investigate the potential of replacing the cost model with LLM itself based on the simplification.

\subsection{Approach}
The essential idea is to let LLM estimate the plan cost given necessary instructions and basic information about the operators and the database. Note that no explicit information is given about how to calculate the cost (like Equation~\ref{eq:cost} or \ref{eq:num-rows}), instead, we directly ask LLM to ``estimate and compare the execution time of the two queries''. In the case of classification, such an approach performs with a high accuracy ranging in 70\%$\sim$90\% on most datasets in our evaluation. And it has the potential to achieve higher accuracy by continuous evolution. Specifically in our task, LLM has great power to summarize the key patterns in the given examples. Considering that real-world query workload usually follows the 80/20 rule (i.e., a large fraction of the queries are similar), LLM can gradually find out the patterns that are key to correctly predicting the execution time for most plans through iterative training. 

The workflow of using LLM as a cost classifier is described as follows:
\begin{enumerate}
    \item \textbf{Initial prompt}: The initial prompt includes basic information about the operators (like definition, format, and the order of single operator costs) as well as simple database statistics including number of rows in each table, number of images in the database and the uniqueness of columns (i.e., whether a column contains duplicate values). In addition, the prompt contains several instructions as external guidance to LLM.   
    \item \textbf{Training}: Given the initial prompt, LLM is first trained to be a classifier. The training data includes several plan pairs and the real execution time for each plan. For the two plans in each pair, we ask LLM for the estimation of execution time and its conclusion on which one is faster. Specifically, the LLM output includes (1) the explanation for the reasoning, (2) the estimated time, and (3) the classification conclusion. Then a judgment on whether its conclusion is correct will be made based on the real execution time. Then the LLM output together with the judgment and ground truth execution time will be appended to the latest prompt as history. Finally, the prompt will consist of the initial prompt and all the history from each training pair. Note that even though we let LLM estimate the time, it is still a classifier rather than a regressor, since we only value the comparison based on the estimation.     
    \item \textbf{Inference}: During inference, the final prompt from the training will be fed into LLM to predict the comparisons on testing plan pairs. No further history will be appended to the prompt at this stage.  
\end{enumerate}

Due to the resource and budget limits, we do not conduct end-to-end experiments where the cost model is replaced with this approach. In the evaluation of this paper, all the experiments except Section~\ref{sec:exp-llm-classifier} use the manually designed cost model. We only evaluate the classification accuracy between this approach and the cost model as an initial exploration for the future research direction. More details are introduced in Section~\ref{sec:exp-llm-classifier}.

\section{Experiments}
\label{sec:exp}
\subsection{Experiment settings}
All experiments are evaluated on a Lambda Quad workstation with 28 3.30GHz Intel Core i9-9940X CPUs, 4 RTX 2080 Ti GPUs, and 128 GB RAM.   

\subsubsection{Evaluation datasets} \hfill\\
\label{sec:eval-ds-and-tasks}
We use four datasets for our evaluation. They are introduced as follows:

\noindent\textbf{ArtBench: }
ArtBench-10 is the first class-balanced, high-quality dataset with clean annotations in the domain of artwork generation datasets. Comprising 60,000 images across 10 distinct artistic styles, 
We use the 256 x 256 version images as the source for our visual operators. The dataset's metadata summarizes diverse information, including image name, artist name, label, length and width, and so on for each artwork, organized into 60,000 rows and 10 columns. 

We leverage the metadata and the images to construct the evaluation database. Specifically, the original metadata is partitioned into three tables, each providing several distinct aspects of information about all the images. One of these tables incorporates paths to corresponding images, such that there is a connection between the two modalities. To ensure the generation of intricate multi-modal queries, all the tables share the key information columns such as artwork ID, artwork name, and artist name, in order to guarantee any two tables are joinable.  

\noindent\textbf{Best-art-of-all-time (BAOAT): } 
The "Best-art-of-all-time" (BAOAT) dataset 
aggregates a comprehensive collection of artworks by the 50 most influential artists in history. 
It includes fundamental details of those artists, such as artists' years of birth and death, genre, nationality, concise biographies, and so on. 
The dataset further provides two versions of images, 
we use the resized version (including 8683 images) as the primary source for our visual-related operators.

The organization of the dataset is originally artist-centric, yielding 50 rows in the metadata corresponding to all 50 artists. 
To augment the intricacy of the metadata structure, we reorganize it to be artwork-centric, where each row furnishes details about an artwork (image), including its ID, name, creator's name, other information of the creator, as well as the file path of this image in the disk. So the new metadata consists of 8683 rows. 
Then we partition the new metadata by column into three distinct tables, sharing a few essential attributes such as artwork ID, artwork name, and artist name to serve for joining. 
Finally, we achieve a more complex multi-modal dataset featuring 8683 images and three tables of metadata each with 8683 rows. 

\noindent\textbf{ART500K: } 
The ART500K dataset represents an expansive visual arts repository, comprising over 500,000 images.
The metadata includes the artist, genre, art movement, event, historical figure, and description for each image. 
We focus on a subset of ART500K, specifically including 43,455 images.

Similar to the datasets above, we partition the metadata into three distinct tables, sharing a few attributes for joining such as artwork ID, artwork name, and artist name. File paths of the images are also included in one table to connect the two modalities. 

\noindent\textbf{Unsplash: } Unsplash is a photography dataset made up of global photos, their metadata (like photo ID and description), and other related data which is sourced from a huge number of user searches (like search keywords). The Unsplash dataset is offered in two versions, lite and full versions, differing in their size. In this paper, we use the lite version, containing 25k nature-themed Unsplash photos, 25k keywords, and 1M searches. It originally has 5 tables with shared keys for joining, so we directly use them without partitioning.

\subsubsection{Evaluation queries}\hfill\\
\label{sec:exp-queries}
We design a random query generator to generate complex and valid initial multi-modal query plans. The maximum number of operators used in each query includes 3 joins, 3 selections, 2 object detections and 2 object countings, and a query includes at least one for each type of the operators. Furthermore, we filter out (1) simple/shallow queries, and (2) significantly time-consuming queries, where the former does not have enough potential to be further optimized, and the latter may take up to several days to execute a single query. The highly similar or near-duplicate queries are also excluded.   
Finally, we collected 39 queries for Artbench, 28 queries for BAOAT, 46 queries for ART500K, and 54 queries for Unsplash.

\subsubsection{Baselines}\hfill\\
\label{sec:baselines}
We select several widely used LLM reasoning methods as baselines. They are introduced as follows:
\begin{enumerate}[leftmargin=2em]
    \item \textbf{Chain of Thought (CoT)}: Following the definition in \cite{wei2022chain}, we provide (1) two demonstration examples for plan optimization; (2) each example contains instances of at least 2 of the 3 policies, each as an optimization step. (3) the explanation for each step (e.g. which operators are moved/merged/removed), given as the optimization thought to guide the reasoning of LLM. 
    
    \item \textbf{Least-to-most}: Similarly to \cite{zhou2022least}, we define partial optimizations of the query as the sub-problems. We first choose two example plans, and for each of them, the whole optimization process is decomposed into several steps, each optimizing a part of the plan (i.e., each corresponding to one sub-problem). 
    Particularly, to make it detailed for Least-to-most to better follow, each step focuses on only one type of operator following only one of the three policies, which results in a significant number of steps that cost non-negligible human effort to make up.      
    In the prompt of the reduction stage, we provide those steps to let LLM learn to decompose the optimization and generate the corresponding instructions by itself. 
    Then in the prompt of problem-solving stage, we ask the language model to optimize the initial plan step by step based on the generated instructions from the previous stage. 
    
    \item \textbf{Self-consistency (SC) enhanced methods}: Following the approach in \cite{wang2022self}, we apply self-consistency (SC) to the baselines above in some experiments. Specifically, the self-consistency mechanism first uses the base method to generate $k$ candidate-optimized plans. 
    Then the most frequent one among the $k$ candidates will be selected as the final output.
    A self-consistency-enhanced method is denoted by \textbf{X-SC} in the rest of this paper, where X is the original method name.
    
    \item \textbf{Aggregation-based methods}: Our challenge lies in constructing/transforming complex structured query plans which are more complicated than the tasks of arithmetic and commonsense reasoning described in \cite{wang2022self}. Consequently, a different reasoning chain usually leads to a different optimized plan. Thus, 
    frequency-based voting of SC is ineffective here, as each candidate is likely unique (in which case the frequency of any candidate is one). Given that the objective of our task is to generate the plan with the shortest execution time, we propose a refined approach based on SC. Specifically, our approach weights each candidate plan by its estimated cost (which approximates the execution time), and we always select the candidate with the lowest cost as the final output when there is a tie on frequency. Particularly, we assign an infinite value as the cost to any invalid plan, i.e., an invalid plan will always have the highest cost to reduce its possibility of being selected. To distinguish our approach and SC, we call our approach \textit{cost-based aggregation}, and for simplification, the term \textit{aggregation} will specifically refer to our cost-based aggregation if there is no explicit statement.
    
    We conduct experiments to compare the performance enhancement achieved by self-consistency (SC) and aggregation in Section~\ref{sec:exp-cost-model}, which shows aggregation improves the base method more significantly. Therefore, we apply aggregation instead of SC to all the evaluated base methods to build their enhanced versions and denote them \textbf{X-agg} where X is the base method name. 
   
\end{enumerate}

\subsubsection{Our methods}\hfill\\

We propose four methods:
\begin{enumerate}
    \item \textbf{\ours}: 
    The fundamental implementation of our method, as depicted in Figures~\ref{fig:arch} and \ref{fig:gcd}, with the full two-level guidance strategy as described in Section~\ref{sec:cost-descent}.

    \item \textbf{\ours-lite}:
    A simplified variant of our method. The only difference is that \ours-lite never provides the cost-feedback (which should include plan cost and whether the plan has been improved) to LLM, i.e., \ours-lite does not have finer-level guidance.
    \item \textbf{\ours-agg}: 
    This variant is implemented by applying aggregation to \ours. 
    \item \textbf{\ours-lite-agg}: 
    This variant is implemented by applying aggregation to \ours-lite. 

\end{enumerate}
The methods with aggregation usually perform more stably than those without it, as the aggregation reduces the randomness in plan generation. Therefore later in the evaluation results, key observations and trends are usually more significant on aggregation-enhanced methods. And if there is no explicit mention, all of our methods use all three policies.

\subsubsection{Implementation}\hfill\\
\label{sec:exp-impl}
To make the evaluation fair and convincing, we develop a lightweight multi-modal query engine to execute the plans, making it straightforward to compare the performance among various methods. The engine is implemented based on our previous work~\cite{dbsim}, with visual operators facilitated by BLIP~\cite{li2022blip}.

\subsubsection{Parameters and metrics}\hfill\\
To sample diverse optimization paths, we follow the setting in \cite{zhou2022least} and apply temperature sampling with T=0.7. When applying aggregation to any base method, we always sample 5 results as the candidates to choose from, i.e., $k=5$. 
With the query engine mentioned above, the performance metrics are (1) the average execution time (in seconds) per plan; (2) time of improvement (ToI): the average improved execution time (i.e., the execution time difference between each initial and optimized plan, in seconds); (3) percentage of improvement (PoI): the average percentage of the reduced time comparing the optimized plan to the initial plan; (4) valid plan ratio (VR): the ratio of valid plans among all the generated plans. 
Note that the execution time only refers to the time of executing the generated plan and excludes the time of the optimization process.

\subsection{End-to-end evaluation}
\label{exp:end-to-end-eval}
In end-to-end evaluation, each initial query plan is optimized respectively by each method. Then both the initial and optimized plans are executed by the query engine mentioned in Section~\ref{sec:exp-impl}. For each query in each dataset, we record the execution time of the initial plan (selected as stated in Section~\ref{sec:exp-queries}) and the optimized plan generated by each method. The execution time is used to calculate the average execution time, PoI, and ToI. Note that the time does not include the optimization time, i.e., here ``execution'' means accepting a given plan (initial or optimized), directly executing it without optimization, and returning the results. Particularly, for an invalid plan, we directly assign its execution time using the time of the corresponding initial plan, i.e., when the optimized plan is invalid, we have no choice but to execute the initial plan to answer the query.           

\begin{figure}
  \centering
  \includegraphics[width=\columnwidth]{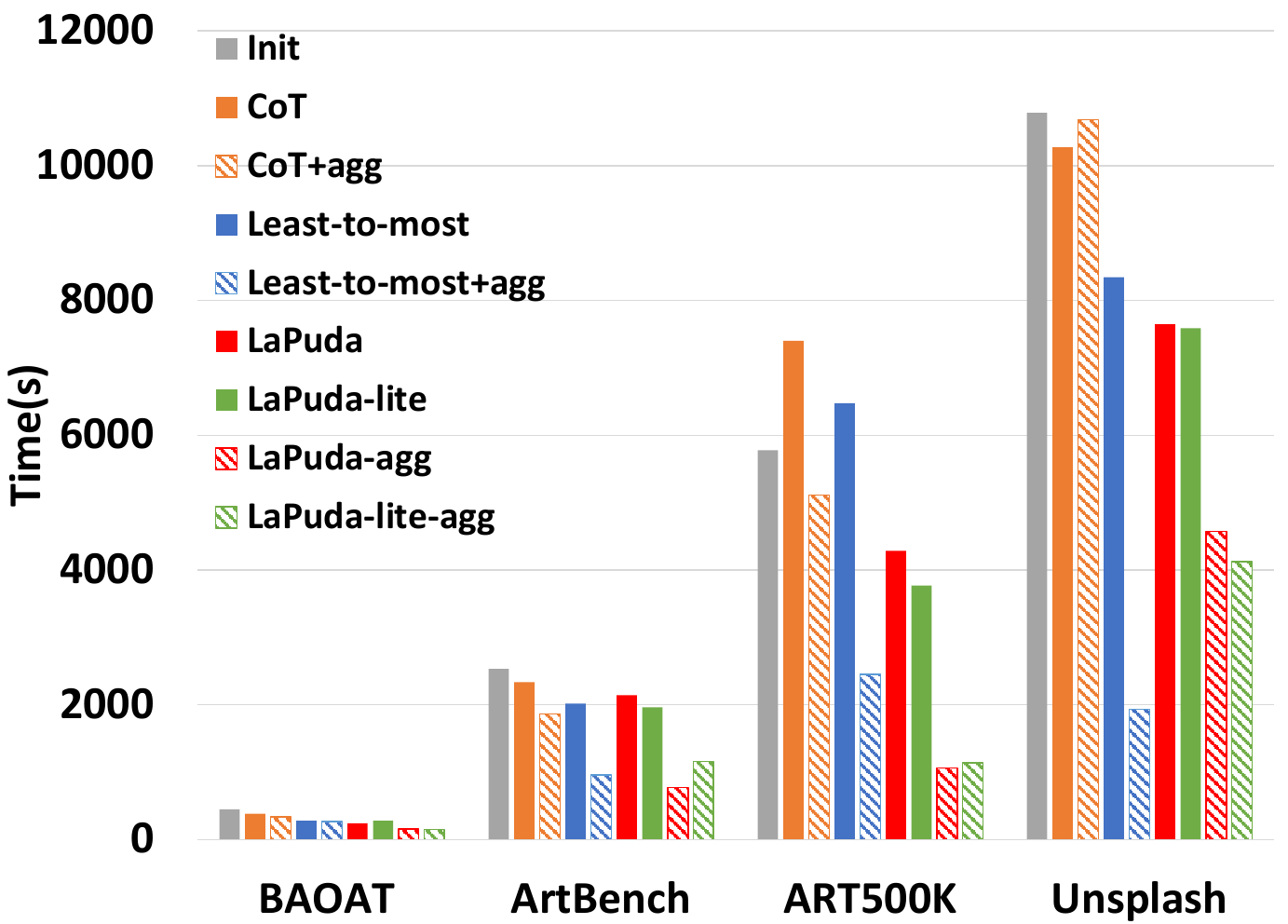}
  \caption{Average execution time of the initial plans and the optimized plans generated by different methods on different datasets}
  \setlength{\belowcaptionskip}{-50pt}
  \label{fig:end2end_exec_time}
  \vspace{-20pt}
\end{figure}

Figure~\ref{fig:end2end_exec_time} illustrates the average execution time per query on different datasets. Specifically, the grey bar presents the average execution time of the initial plans for each dataset, while other bars show the average execution time of the optimized plans generated by each method. Particularly, a solid bar (except the initial plan bar) represents a method without aggregation, and its aggregation version is shown as the backslash-filled bar with the same color. 
In most datasets, \ours\ (red solid bar) and \ours-lite (green solid bar) achieve higher optimization quality than the baselines without aggregation, i.e., the average execution time of the plans optimized by \ours\ and \ours-lite is usually much less than that of Least-to-most (blue solid bar) and CoT (orange solid bar). The aggregation methods demonstrate a similar trend more significantly: the plans generated by \ours-agg and \ours-lite are executed 1$\sim$3x faster than those of CoT-agg and up to 1x faster than Least-to-most-agg on most datasets. Furthermore, \ours-agg and \ours-lite-agg generate the overall most efficient optimized plans for 3 out of the 4 datasets (except Unsplash).

We observe that sometimes the non-aggregated baselines will generate negative optimization, i.e., execution of optimized plans is overall slower than the initial plans. For example, the generated plans of Least-to-most and CoT have longer execution time than the initial plans (which is also reflected by the negative PoI and ToI in Table~\ref{table:main-result}) on the ART500K dataset. This is because they are not aware of the plan cost, making it hard for them to determine whether the current optimization goes towards the optimal, thus those methods may keep going on wrong direction and result in negative optimization. And their aggregation versions do not have such a problem, which proves the importance of involving cost information in the optimization. Moreover, \ours{} and \ours-lite still perform well on this dataset, further affirming the strength of our methods.      

In addition to the absolute end-to-end execution time of plans in Figure~\ref{fig:end2end_exec_time}, we report more detailed results about the optimization quality in Table~\ref{table:main-result}, including (1) improvement on the plan execution efficiency, i.e., the execution time difference (both absolutely and relatively) between the initial and optimized plans, and (2) the capability of each method to avoid plan errors, i.e., the ratio of valid plans.
A consistent trend across all datasets is the superior performance of \ours\ with aggregation (\ours-agg) in both PoI (Percentage of Improvement) and ToI (Time of Improvement). This underscores the outstanding effectiveness and robustness of our methods. In addition, its aggregated variant (\ours-lite-agg) which lacks cost-feedback, also achieves competitive performance on all datasets, which further justifies the use of the cost model and policies in our methods.  
For example, on ART500K dataset, \ours-agg outperforms other aggregated methods with up to 0.43 difference on PoI and at most around 4000s improvement on ToI. \ours{} also has a higher performance than other non-aggregated methods. Similar trends are displayed on other datasets, too.   
Besides, all methods with aggregation significantly outperform their non-aggregated counterparts in terms of all three matrices (PoI, ToI and VR), underscoring the value of our proposed cost-based aggregation in enhancing query optimization on both execution speed and correctness concerns. 

When we consider the ability to generate valid optimized plans, Least-to-most and its derivatives exhibit high consistency across all datasets, \ours{} and \ours-lite achieve competitive VR against Least-to-most, while CoT always performs the worst on VR (i.e., it generates many invalid plans). This indicates the excellent reliability of our methods in producing correct optimized plans.

On Unsplash, an interesting phenomenon is the occasional inconsistency between PoI and ToI, i.e., some method excelling in PoI may has a relatively lower ToI than another method, or vice versa. For example, Least-to-most-agg has a lower PoI but higher ToI compared to \ours-agg.     
Such an inconsistency happens when one method performs better on a few time-consuming queries while another method makes better optimization for other queries. In such a situation, though the former method achieves more improvement on the absolute execution time (which results in higher ToI) of those time-consuming queries, the relative improvement of time (PoI) is not significant since the denominator is large. But overall, this phenomenon happens infrequently and the PoI and ToI are still consistent in most cases.

In summary, across all datasets, \ours{} and its variants consistently outperform the baselines, which confirms the suitability of our methods for multi-modal query optimization tasks.

\begin{table*}[]
    \begin{tabular}{@{}l|ccc|ccc|ccc|ccc@{}}
        \toprule
        & \multicolumn{3}{c|}{BAOAT}    & \multicolumn{3}{c|}{ArtBench} & \multicolumn{3}{c|}{ART500K} & \multicolumn{3}{c}{Unsplash}  \\ \midrule
         & PoI $\uparrow$ & ToI $\uparrow$ & VR $\uparrow$ & PoI $\uparrow$ & ToI $\uparrow$ & VR $\uparrow$ & PoI $\uparrow$ & ToI $\uparrow$ & VR $\uparrow$ & PoI $\uparrow$ & ToI $\uparrow$ & VR $\uparrow$\\ \midrule
    CoT                         & 0.14    & 64.04    & 0.33    & 0.10    & 195.93  & 0.36    & -0.59   & -1623.16 & 0.16   & 0.03    & 509.61  & 0.04 \\
    CoT-agg              & 0.29    & 108.76   & 0.45    & 0.34    & 670.16  & 0.64    & 0.32    & 666.30          & 0.64   & 0.08    & 99.12   & 0.16 \\
    Least-to-most               & 0.37    & 169.49   & \textbf{1.00}     & 0.14    & 513.34   & 0.96    & -0.31          & -695.58 & \textbf{1.00}    & 0.30     & 2436.98 & 0.82  \\
    Least-to-most-agg    & 0.68    & 293.27   & \textbf{1.00}     & 0.63    & 1575.26  & \textbf{1.00}  & 0.55   & 3328.50  & \textbf{1.00}   & 0.41     & \textbf{8848.18} & \textbf{1.00}  \\ \midrule
    \ours                       & 0.48    & 205.50   & 0.78    & 0.20    & 391.30  & 0.80     & 0.47    & 1493.00  & 0.84   & 0.23    & 3135.92 & 0.73  \\
    \ours-lite                  & 0.33    & 168.50   & 0.89    & 0.18    & 569.32  & 0.88    & 0.47    & 2008.57  & 0.88   & 0.29    & 3195.80 & 0.61 \\
    \ours-agg                    & \textbf{0.69}      & 289.09  & 0.89    & \textbf{0.70}   & \textbf{1761.70}  & 0.96   & \textbf{0.75}    & \textbf{4716.36}   & 0.92    & \textbf{0.51}    &  6207.27 & 0.88    \\ 
    \ours-lite-agg               & 0.67    & \textbf{297.90}    & 0.96    & 0.52    & 1375.91 & \textbf{1.00}    & 0.74   & 4636.87          & \textbf{1.00}      & 0.31    & 6657.41 & 0.92 \\
    \bottomrule
    \end{tabular}
     \caption{End-to-end comparison of all methods on different datasets. PoI, ToI and VR refer to the average percentage of running time improvement, the average absolute running time improvement, and the ratio of valid plans in the generated plans by each method.}
     \label{table:main-result}
     \vspace{-15pt}
\end{table*}

\begin{table*}[ht!]
    \centering
    \begin{tabular}{c c c c c c c c c}
        \toprule
        \textbf{Dataset} &  & \textbf{LaPuda-No Policy} & \textbf{LaPuda-Policy 1} & \textbf{LaPuda-Policy 2} & \textbf{LaPuda-Policy 1+2} & \textbf{LaPuda (3 Policy)}\\
        \midrule
        BAOAT & PoI $\uparrow$ & 0.31 & 0.40 & 0.20 & 0.51 & 0.48                   \\
        & ToI (s) $\uparrow$ & 137.60 & 169.90 & 100.00 & 220.54 & 205.50           \\
        & VR $\uparrow$ & 0.56 & 0.78 & 0.78 & 0.78 & 0.78                          \\
        \midrule
        ArtBench & PoI $\uparrow$ & 0.19 & 0.25 & 0.17 & 0.32 & 0.20                \\
        & ToI (s) $\uparrow$ & 414.73 & 529.87 & 345.01 & 743.31 & 391.30           \\
        & VR $\uparrow$ & 0.64 & 0.80 & 0.84 & 0.84 & 0.80                          \\
        \midrule
        ART500K & PoI $\uparrow$ & 0.25 & -0.81 & 0.33 & 0.52 & 0.47                \\
        & ToI (s) $\uparrow$ & 765.83 & -2496.67 & 1125.10 & 2535.52 & 1493.00      \\
        & VR $\uparrow$ & 0.52 & 0.60 & 0.68 & 0.88 & 0.84                          \\
        \midrule
        Unsplash & PoI $\uparrow$ & 0.19 & 0.14  & 0.15 & 0.04 & 0.23               \\
        & ToI (s) $\uparrow$ & 1926.82 & 2937.78 & 3148.82 & 2338.55 & 3135.92      \\
        & VR $\uparrow$ & 0.35 & 0.61 & 0.53 & 0.61 & 0.73                          \\
        \bottomrule
    \end{tabular}
    \caption{Evaluation of different policies}
    \label{tab:policy}
    \vspace{-20pt}
\end{table*}

\subsection{Impact of policies}
\label{sec:exp-policy}
To evaluate the distinct effect of different policies (introduced in Section~\ref{sec:Policy}) in our methods, we conduct a series of experiments to reveal the diverse impact of individual policies and their combinations on the performance of \ours. 
The results are reported in Table~\ref{tab:policy}, where we use ``\ours-No Policy'', ``\ours-Policy 1'', ``\ours-Policy 2'', ``\ours-Policy 1+2'', ``\ours{}(3 Policy)'' to respectively denote \ours\ without any policies, \ours\ with only policy 1 or 2, \ours\ with both policies 1 and 2, as well as the full \ours. Here \ours-No Policy is used as the baseline.      

On BAOAT, ArtBench, and ART500K datasets, compared to the no-policy baseline, the individual Policy 1 and 2 always show opposite impact on the performance of \ours, i.e., one improves PoI and ToI while another downgrades them. For example, on BAOAT, \ours-Policy 1 achieves higher PoI and ToI than \ours-No Policy while those of \ours-Policy 2 are lower than \ours-No Policy.
But combining Policy 1 with Policy 2 can further improve the performance of the better one (i.e., \ours-Policy 1 on BAOAT and ArtBench, \ours-Policy 2 on ART500K). It is because (1) moving operators (Policy 1) enables more operators to be merged (Policy 2), since merging only occurs over neighboring operators, and (2) merging operators eases the movement as the structure complexity of the plan is reduced. Therefore, the combination of the two policies outperforms the individual policies.      

Another interesting observation is that on those three datasets, the combination of Policies 1 and 2 outperforms that of all three policies. By further analyzing the optimization procedure, we notice that LLM encounters challenges in understanding and applying Policy 3 (operator removal), which is partially reflected in the slightly lower VR of full \ours{} than \ours-Policy 1+2 in many cases (like on ArtBench and ART500K datasets).     
The reason is possibly that Policy 3 is more open and complex than 1 and 2. Specifically, Policy 1 only considers plan structure, Policy 2 cares about the semantics between same-type operators (e.g., merging two selections, merging two object detections, etc.), while Policy 3 additionally focuses on the semantics between operators of different types, like object counting and detection, resulting in more possible optimization directions. Therefore, LLM may not be able to fully understand Policy 3 in some cases, and the misunderstanding is likely to negatively affect the overall performance.
Although Policy 3 slightly downgrades the method performance on some datasets, it still produces a positive impact and significantly improves the performance on other datasets. For example, on the Unsplash dataset, compared to \ours-Policy 1+2, full \ours{} achieves around 6x PoI and 1.3x ToI.
Besides, given that the operator removal policy covers a big family of commonly used rules in query optimization, Policy 3 remains a critical component of our methods, and we still enable Policy 3 in our methods for end-to-end evaluation.

Furthermore, across all datasets, a general trend is that the application of policies improves VR compared to \ours-No Policy. This remarks on the effectiveness of our proposed policies in guiding LLM to guarantee the generated plan's correctness.  

\subsection{Aggregation vs. self-consistency}
\label{sec:exp-cost-model}
As mentioned in Sec~\ref{sec:baselines}, we adjust the self-consistency reasoning (which is originally frequency-based as stated in Section~\ref{sec:baselines}) to voting by cost (i.e., the \textit{aggregation}) for our query optimization task. To demonstrate the effectiveness of aggregation, we compare the optimization quality between a non-aggregated method + self-consistency (SC) vs. the same method + aggregation. 
Due to budget limits, we conduct the evaluation on all datasets using CoT as the only base method. 
Respective results are presented in Table~\ref{tab:Agg}. 

The results indicate that aggregation yields an evident increase in PoI compared to self-consistency across all datasets, as well as significantly higher ToI on most datasets. This suggests that our cost-based aggregation is more reasonable and effective than the original self-consistency in the query optimization scenario. 
Notably, on Unsplash, the ToI demonstrates a reduction from CoT-SC to CoT-agg, with an increase in PoI. Such an inconsistency between PoI and ToI is probably due to the same reason as in Section~\ref{exp:end-to-end-eval}. 

Furthermore, CoT-agg leads to a pronounced improvement in the generated plan correctness, reflected by the non-trivial increase of VR. Specifically, CoT-agg is aware of the cost, making the invalid plans less likely to be selected as the final output, as they are always associated with the highest cost (mentioned in Section~\ref{sec:baselines}). 

\begin{table}[ht!]
    \centering
    \begin{tabular}{c c c c }
        \toprule
        \textbf{Dataset} &  & \textbf{CoT-SC } & \textbf{CoT-agg } \\
        \midrule
        BAOAT & PoI$\uparrow$ & 0.14 & 0.29     \\
        & ToI (s)$\uparrow$ & 57.61 & 108.76    \\
        & VR$\uparrow$ & 0.33 & 0.45            \\
        \midrule
        ArtBench & PoI$\uparrow$ & 0.02 & 0.34  \\
        & ToI (s)$\uparrow$ & 172.37 & 670.16   \\
        & VR$\uparrow$ & 0.38 & 0.64            \\
        \midrule
        ART500K & PoI$\uparrow$ & -0.77 & 0.32  \\
        & ToI (s)$\uparrow$ & -1222.44 & 666.30 \\
        & VR$\uparrow$ & 0.50 & 0.64            \\
        \midrule
        Unsplash & PoI$\uparrow$ & 0.01 & 0.08  \\
        & ToI (s)$\uparrow$ & 276.32 & 99.12    \\
        & VR$\uparrow$ & 0.02 & 0.16            \\
        \bottomrule
    \end{tabular}
    \caption{Evaluation of aggregation}
    \label{tab:Agg}
    \vspace{-20pt}
\end{table}

\subsection{Impact of finer-level guidance}
The experiment results in Table~\ref{table:main-result} present the effectiveness of the finer-level guidance (cost-feedback) in \ours. Compared to \ours-lite-agg, \ours-agg achieves significantly higher PoI and ToI in most cases, showing that the finer-level guidance does have a non-trivial impact on improving the optimization quality. Without aggregation, \ours\ does not always have a higher ToI than \ours-lite in Table~\ref{table:main-result}, but the PoI of \ours\ is still higher in most cases. 

We also observe that the ratio of valid plans (VR) of \ours\ is usually lower than \ours-lite, and the aggregation versions also show a similar trend. This is probably because the cost-feedback introduces not only more information but also more noise, as the cost estimation unavoidably includes errors. However, even having this issue, \ours\ still overall outperforms \ours-lite (with or without aggregation), which further proves the power of the two-level guidance strategy.

\subsection{Performance of LLM as cost model}
\label{sec:exp-llm-classifier}
In this section, we evaluate the prediction accuracy of the LLM-based execution time classifier vs. our cost model, as introduced in Section~\ref{sec:llm-cost-model}. We select the optimized plans generated by \ours\ and Least-to-most to construct the testing plan pairs, i.e., for each query, the corresponding plan pair is ($p_{lp}$, $p_{lm}$) where $p_{lp}$ is the optimized plan by \ours\ and $p_{lm}$ is generated by Least-to-most. For each dataset, we filter out the queries for which the plan pair includes invalid plan(s ), i.e., if $p_{lp}$ or $p_{lm}$ for a query is invalid, the testing will exclude that query. After the filtering, we have 22, 29, 36, 32 queries for BAOAT, ArtBench, AR500K, and Unsplash respectively. Then for each dataset, we split the filtered queries into training and testing sets by a ratio of 5:5. Then LLM classifier is trained on the training set as in Section~\ref{sec:llm-cost-model} and both the LLM classifier and cost model are tested on the testing set, i.e, the accuracy results reported in Table~\ref{tab:llm-cost-model} are computed on the testing set.

\begin{table}[h]
    \centering
    \begin{tabular}{c c c }
        \toprule
        \textbf{Dataset} & \textbf{LLM-classifier} & \textbf{Cost-model} \\
        \midrule
        BAOAT & \textbf{0.75} &  0.67    \\
        \midrule
        ArtBench & 0.87 & \textbf{1.0}  \\
        \midrule
        ART500K & 0.56 & \textbf{0.72}  \\
        \midrule
        Unsplash & 0.69 & \textbf{0.75}  \\
        \bottomrule
    \end{tabular}
    \caption{Prediction accuracy of LLM classifier and cost model on execution time comparison}
    \label{tab:llm-cost-model}
    \vspace{-20pt}
\end{table}
 
The results show that the LLM cost classifier performs better than the cost model on BAOAT while the cost model wins on other datasets. However, it does not deny the potential of the LLM classifier. As we mentioned in Section~\ref{sec:llm-cost-model}, the investigation for the LLM classifier is just in the early stage, and the LLM classifier has the potential to further improve its performance given more training data. Considering that it has achieved around 70\%$\sim$90\% accuracy on at least three datasets, it is reasonable that we can advance the LLM cost classifier to be as effective as the cost model in query optimization given better instructions and more training queries, which will place the last piece of the puzzle on designing an end-to-end query optimizer purely using LLM.

\section{Conclusion}
In this paper, we propose \ours, an LLM and policy-based
multi-modal query optimizer with novel guided cost descent (GCD) and two-level guidance mechanism. Experiments present that \ours\ outperforms the state-of-the-art LLM reasoning methods on multi-modal query optimization tasks: \ours\  achieves significantly higher optimization effectiveness (which results in faster execution of the optimized plans) than the baselines. \ours\ is the first step towards a brand new design pattern that builds a query optimizer centered on LLM, which can serve as the next-generation query optimizer in the era of multi-modality.  

\bibliographystyle{ACM-Reference-Format}
\bibliography{main}

\end{document}